\documentstyle[epsfig,epsf,art12,fullpage]{article}
\begin{document}
\par

\newpage

\begin{center}
{\large V.Ammosov, V.Gapienko, A.Ivanilov, A.Semak,\\[0.5cm]
Yu.Sviridov$^*$, E.Usenko, V.Zaets}\\[1cm]
\end{center}

\begin{center}
{\it State Research Centre of Russian Federation\\[0.5cm]
Institute for High Energy Physics\\[0.5cm]
142281 Protvino Moscow region Russia}\\[2cm]
\end {center}

\begin{center}
{\Large \bf COMPARISON OF TIMING PROPERTIES\\[0.5cm]
OF GLASS MULTIGAP RPCs\\[0.7cm]
WITH 0.3 AND 0.6 MM SUBGAP WIDTH}\\[3cm]
\end{center}

\begin{center}
{\large \bf Abstract}
\end{center}
 
Two glass RPCs with symmetrical read-out scheme --- (2$\times$(2$\times$0.3mm)) and 
(2$\times$0.6mm) --- were constructed and tested at CERN PS T10 beam. 
Specially developed electronics was used.
 Electronics resolution convoluted with measured charge spectra at operating point 
was estimated as about 65 ps. 
 The best time resolutions obtained 
at low counting rate were 75 ps for the 0.3 mm subgap and about 125 ps 
for the 0.6 mm one, after slewing corrections.  
 The rate capability of the (2$\times$0.6mm) counter 
was found to be much worse than for the (2$\times$(2$\times$0.3mm)) one. 
   
\vfill



\vfill
 
$^{*)}$ Corresponding author, E-mail: yu\_sviridov@mx.ihep.su

\newpage

\begin{center}
{\Large \bf INTRODUCTION}
\end{center}

In the recent years significant progress was achieved in developing 
 RPCs working in avalanche mode at
 atmospheric pressure, as TOF detectors [1].  
From the point of view of counters design,
 the following direction was formulated: 

---using narrow (sub-millimeter) gas gaps to obtain excellent time resolution;

---summing signals of several such subgaps to have high efficiency, 
good charge spectrum and to reject tail of delayed signals. 

Multigap counters having 5---6 subgaps as narrow as 0.21 mm 
are under investigation as candidates for the ALICE TOF system [2]. 
Other groups are developing 4-gaps counters with subgap width 0.3 mm [3].   

But up to now there is not enough information about the dependence 
 of RPC time resolution 
 on gas gap width. 
Time resolution of about 70 ps was obtained
 for (5$\times$0.22 mm) multigap glass 
chamber [2] and about 50 ps --- for the (2$\times$(2$\times$0.3 mm)) small 
counters [3].
In [4], about 100 ps resolution was obtained for the four-gaps counter 
with 0.3 mm gap width; for the double-gap counter with 0.6 mm wide 
subgaps, the measured time resolution was 180-200ps [5]. 

The only reserved study of the time resolution dependence 
on gas gap width was made in [6].
The time resolutions of about 120, 170 and 230 ps were measured 
for the 0.3, 0.4 and 0.6 mm gas gaps, respectively. 
 It was found, that for this  
range of  gap width, time resolution can be presented as
 a linear function 
of gap width with the slope of 40 ps/100 $\mu$m.  

In the course of our work on the HARP experiment [7] TOF system we have 
studied two options of glass RPCs, one having 4 subgaps 0.3 mm wide 
and the other --- 2 subgaps 0.6 mm wide.
The construction of counters having less 
number of wider subgaps is obviously  
simpler and less expensive. We studied what will be the retribution 
of this approach in terms of operational characteristics of the counter. 
Though the decision in HARP was made in favour of the 0.3 mm wide gap [8],
 we believe 
that our results have nevertheless more general interest.    

\begin{center}
{\Large \bf 1. COUNTERS DESIGN AND FRONT END ELECTRONICS}
\end{center}

Basing on our previous experience, we used as most promising the 
symmetrical read-out scheme. That is, in case of 0.6 mm gap,
 the read-out electrode was
 placed between two such subgaps. In case of 0.3 mm gap, read-out
electrode was placed between two (2$\times$0.3 mm) double-gaps, 
thus forming (2$\times$(2$\times$0.3 mm)) structure.
This approach provides better read-out efficiency (the ratio 
between measured charge and the avalanche charge inside gas gap) and  
requires lower voltage than fully serial multigap counters. 
  
 As it is seen, 
in both cases the total gas width was equal to 1.2 mm, to have high 
efficiency (we have found in course of preparatory studies, that 
1 mm total gas thickness gives about 2\% lower efficiency). 

Glass electrodes were made of 1 mm thick window glass with sensitive 
area of 130$\times$200mm$^2$. Glass volume resistivity was 
measured to be about 7$\times$10$^{12}$ $\Omega$cm. Uniform gap width
 was provided by fishing lines with 0.3 and 0.6 mm diameters, 
respectively, placed with $\simeq$40 mm step between glass plates and 
fixed with epoxy drops in few points. High voltage was distributed on external 
surfaces of each substructure (0.6 mm subgap or (2$\times$0.3 mm) 
double-gap) using carbon film. Assemled glass stacks were put into gas tight 
aluminum boxes. 

We used gas mixture C$_2$H$_2$F$_4$/iC$_4$H$_{10}$/SF$_6$ in flow rate 
composition 90/5/5. 

Specially developed electronics consisted (Fig. 1) of the on-counter 
preamplifier (PA) and remote Splitter-Shaper-Disriminator (SSD) in 
CAMAC standard. The PA had 100 mV/pC conversion factor and 2 GGz bandwidth. 
PA output was fed to SSD by coaxial cable about 0.5 m long. SSD provided further 
signal amplification by 2 and splitted it in two branches. One branch --- 
Shaper --- lengthened the tail of the short input pulse for better transition 
to the remote QDC. The other branch was the discriminator with variable 
(4 --- 10 mV) threshold; output NIM signal of the discriminator was sent 
to TDC. Measured FEE resolution in dependence on input charge is shown on 
Fig. 2. 

Both detectors were tested at T10 CERN PS beam of 7 GeV/c pions using 
ALICE TOF group environment and DAQ. Trigger counters defined 1$\times$1 
cm$^2$ area on the RPCs. Time resolution of RPCs was obtained by measuring 
$\Delta$T between RPC signal and the time mark, provided by scintillation 
counter with 40 ps resolution, also written to TDC. In the results 
presented further, the resolution of start counter is quadratically 
subtracted.  

\begin{center}
{\Large \bf 2. EXPERIMENTAL RESULTS}
\end{center}

Time resolution of RPCs was obtained off-line after slewing correction. 
The examples of
 charge spectra, arrival time - charge correlation plots and corrected 
$\Delta$T distributions for both counters are shown on Figures 3 and 4.
 Time - charge correlation plots were fitted with appropriate polinom 
and the corrected time distributions were obtained with respect to this 
fitting function. These $\Delta$T distributions were fitted with 
Gaussian, and the standard deviations of the obtained fit are 
refered  further on as time resolutions.  

The results are shown on Fig. 5 as a function of applied HV.
Discrimination threshold was set at 4 mV.
 Total efficiency for the both counters was about 99\% in operating region. 
 For both counters, 
two read-out pads were studied differing in area ten times. For the small 
pads, the best measured time resolutions are 75 ps for the 0.3 mm subgap and
 about 140 ps for the
 0.6 mm one. These resolutions were achieved at 6.2 and 5.4 kV, respectively.        
 Data for the 0.3 mm counter were measured at beam intensity less 
than 60 Hz/cm$^2$ during spill of 300 ms duration. Results for the 
0.6 mm counter presented on this Figure were obtained at beam intensity 
of about 140 Hz/cm$^2$. As it will be seen later, the best measured 
time resolution for the (2$\times$0.6 mm) counter at counting rate 
lower than 100 Hz/cm$^2$, was 125 ps. 

For both counters, ten times larger pads show worse resolution; 
the difference is about 17 ps for the 0.3 mm subgap and 25 ps --- 
for the 0.6 mm one. The best time resolution was obtained at about 
1 pC mean charges for both counters. 

Very significant difference was observed in the rate dependence of $\sigma$$_t$ 
for two counters, Fig. 6. The data refer to different pads: it was 100 cm$^2$ 
for 0.3 mm subgap and 9 cm$^2$ for the 0.6 mm one. Degradation of resolution is 
dramatic for the (2$\times$0.6 mm) counter; there is practically no region 
of constant time resolution. On the contrary, the (2$\times$(2$\times$0.3 mm)) 
counter shows much slower resolution degradation. For the rate range up to 
2 kHz/cm$^2$, width of the time distribution for the 0.3 mm subgap 
is increased on only 50 ps, while this rise is about 200 ps for the 0.6 mm 
subgap. 

On Fig. 7, the dependence of mean measured charges on particle rate 
is shown for both counters. 
 Average charge for the (2$\times$0.6mm)
counter is decreased about 5 times for the 
rate range from 0.1 to 2 kHz/cm$^2$. For the 0.3 mm subgap counter,
 this dependence is 
considerably  
weaker. Such behaviour can be explained by the drop of effective 
potential difference applied to gas gap due to increased current
 flow through  
glass electrodes of high resistivity. 
 Accordingly the counting rate has two-fold 
influence on the time resolution: decrease in effective electric field 
strength E in gas gap and decrease in amplitudes of signals. 
Fig. 7 shows that this effect is more important for the double-gap 
counter than for the four-gaps one.       

\begin{center}
{\Large \bf 3. DISCUSSION}
\end{center}

As it was said, the best measured time resolutions were 75 ps for the 0.3 mm 
subgap width and 125 ps --- for the 0.6 mm one. Some remarks should be made 
however. 

Keeping in mind very strong dependence of time resolution on the particle
 rate for the 0.6 mm subgap counter, we can extrapolate measured dependence
 (Fig. 6) to obtaine $\sigma$$_t$ $\simeq$ 100 ps for this detector at low rate. 
 FEE electronics resolution averaged over measured charge 
spectra (Fig's 3 and 4) can be estimated approximately as 65 ps.
 After quadratic subtraction of this estimate from the above values,     
 one obtaines resolutions of 40 ps and 80 ps
 for the 0.3 mm and 0.6 mm subgaps, respectively. 

What can be expected for $\sigma$$_t$ vs gap width dependence? 
It is known that time resolution is proportional to the rise time of the 
detector signal T$_f$, which in turn in our case is proportional 
to 1/$\alpha$v$_d$. Here $\alpha$ is effective first Townsend coefficient 
 and v$_d$ --- electrons drift velocity for the gas mixture used.  

Both these quantities are expected to increase
 with electrical field strength E, 
which was 10.3 kV/mm for 0.3 mm counter and 8.75 kV/mm for the 0.6 mm one, 
at points of the best resolution. There are 
unfortunately no measurments
 of $\alpha$ and v$_d$ in this region for the gas mixture used.  
Quite recently [9] the results of simulation of these quantities 
in dependence on E were reported. According to these data, 
the product $\alpha$v$_d$ decreases 1.7 times for the E range 
from 10.3 kV/mm to 8.75 kV/mm, which seems to be compatible with our 
estimate $\sigma$$_{0.6}$/$\sigma$$_{0.3}$$\simeq$2, taking into account 
uncertanties in our analysis.

\begin{center}
{\Large \bf CONCLUSION}
\end{center}

Timing characteristics of two multigap Resistive Plate Counters with different 
subgap width of 0.3 mm and 0.6 mm were measured. The same specially developed 
FEE 
 was used.

The best measured time resolutions were 75 ps and 125 ps for the
 (2$\times$(2$\times$0.3mm)) and (2$\times$0.6mm)  
counters, respectively. It was found that the later counter is much more  
sensitive to the particle rate than the former one. 

Our results indicate also that the time resolutions achievable 
for the two counters are about 40 and 80 ps for the 0.3 and 0.6 mm 
subgaps, respectively.

\begin{center}
{\Large \bf ACKNOLEDGEMENTS}
\end{center}

The authors would like to thank M.C.S.Williams and ALICE TOF group 
for continuous support.    

\newpage

\begin{center}
{\Large \bf REFERENCES}
\end{center}

\begin{itemize}
\item[1.] W.Klempt, Nucl. Instr. Meth., Vol. A433, 1999, 542.
\item[2.] A.Akindinov et al., Nucl. Instr. Meth., Vol. A456, 2000, 16.
\item[3.] P.Fonte et al., Nucl. Instr. Meth., Vol. A449, 2000, 295.
\item[4.] A.Akindinov et al., ITEP preprint 20-99, Moscow, 1999. 
\item[5.] A.Akindinov et al., ITEP preprint 45-98, Moscow, 1998.
\item[6.] P.Fonte et al., CERN preprint CERN-EP/99-68, Geneva, 1999. 
\item[7.] HARP proposal, http://harp.web.cern.ch/harp.
\item[8.] M.Bogomilov et al., The RPC 
time-of-flight system of the HARP experiment, presented by J.Wotschack at the
 6th Int. Workshop on RPCs and 
Related Detectors, Coimbra, Portugal, 2001;
 http://doc.cern.ch/DTP/dtp-2001-029/files/MainFile/.
\item[9.] Talk of W.Riegler at the  
6th Int. Workshop on Resistive Plate Chambers and Related Detectors,
Coimbra, Portugal, 2001;http://www-lip.fis.uc.pt/\~~rpc2001/pdf/riegler.pdf.  
\end{itemize}

\newpage

\begin{figure}[hbt] 
\vbox to15cm{
\vfill
\includegraphics{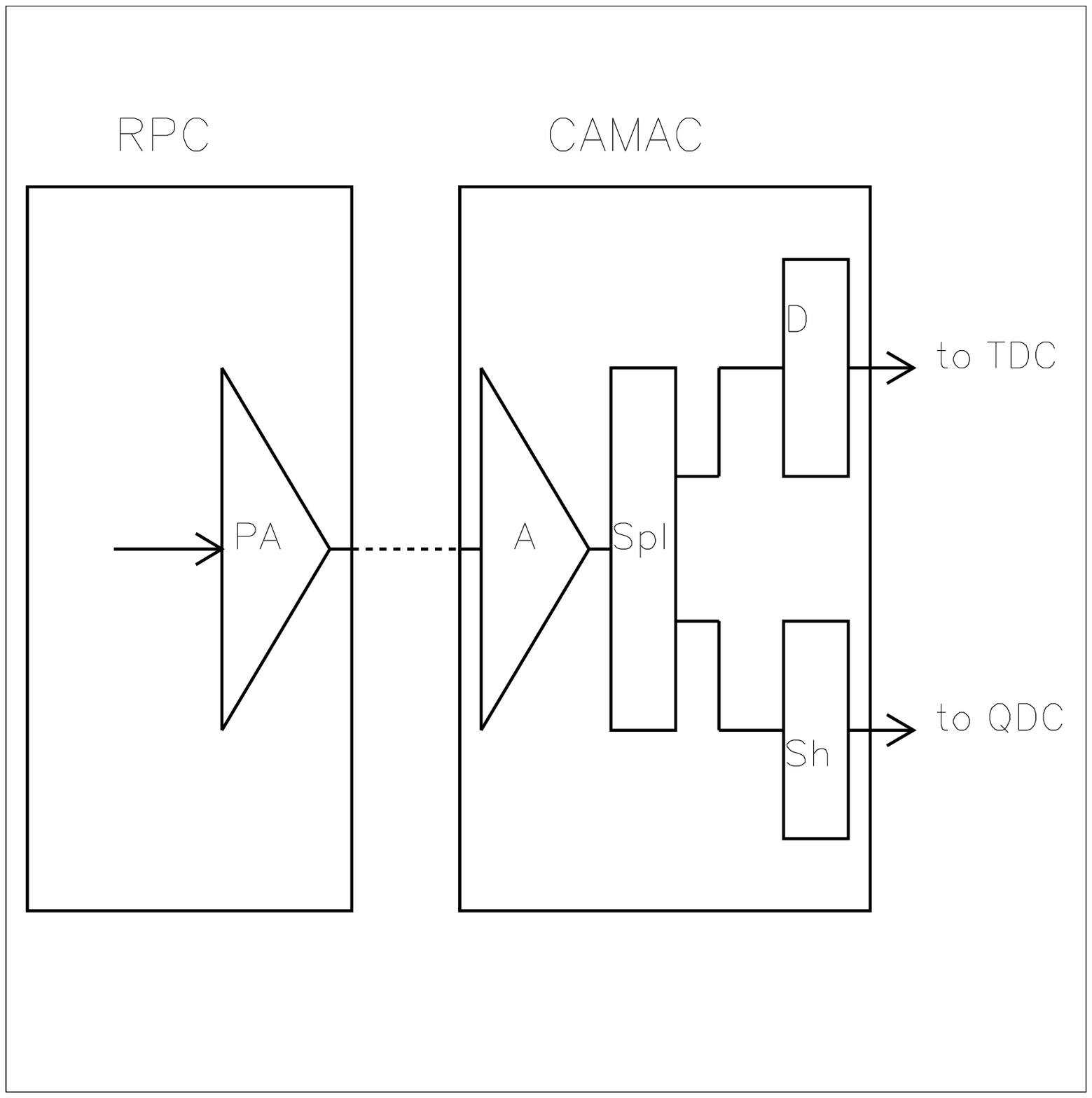}
\vfill
}
\vskip 7.5cm
\begin{minipage}[t]{16cm}
\begin{center}
\bf{Fig. 1.} 
\hskip 0.5cm
{Block-scheme of electronics: PA --- preamplifier, 
A --- amplifier, Spl --- splitter, D --- discriminator, Sh --- analog shaper.} 
\end{center}
\end{minipage}
\end{figure}

\begin{figure}[hbt] 
\vbox to15cm{
\vfill
\includegraphics{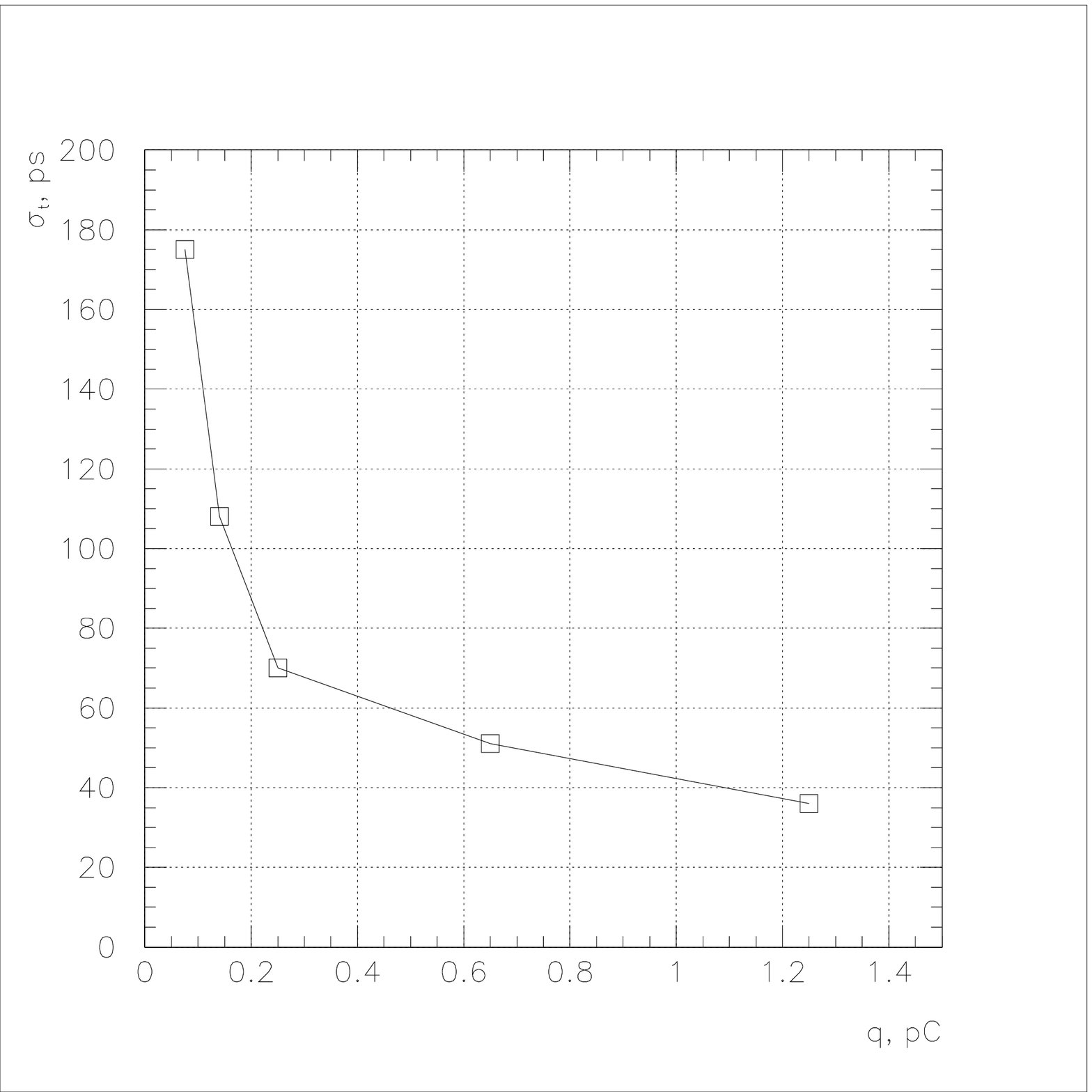}
\vfill
}
\vskip 10.5cm
\begin{minipage}[t]{17cm}
\begin{center}
\bf{Fig. 2.} 
\hskip 0.5cm
{FEE resolution in dependence on input charge.}  
\end{center}
\end{minipage}
\end{figure}

\begin{figure}[hbt] 
\vbox to15cm{
\vfill
\includegraphics{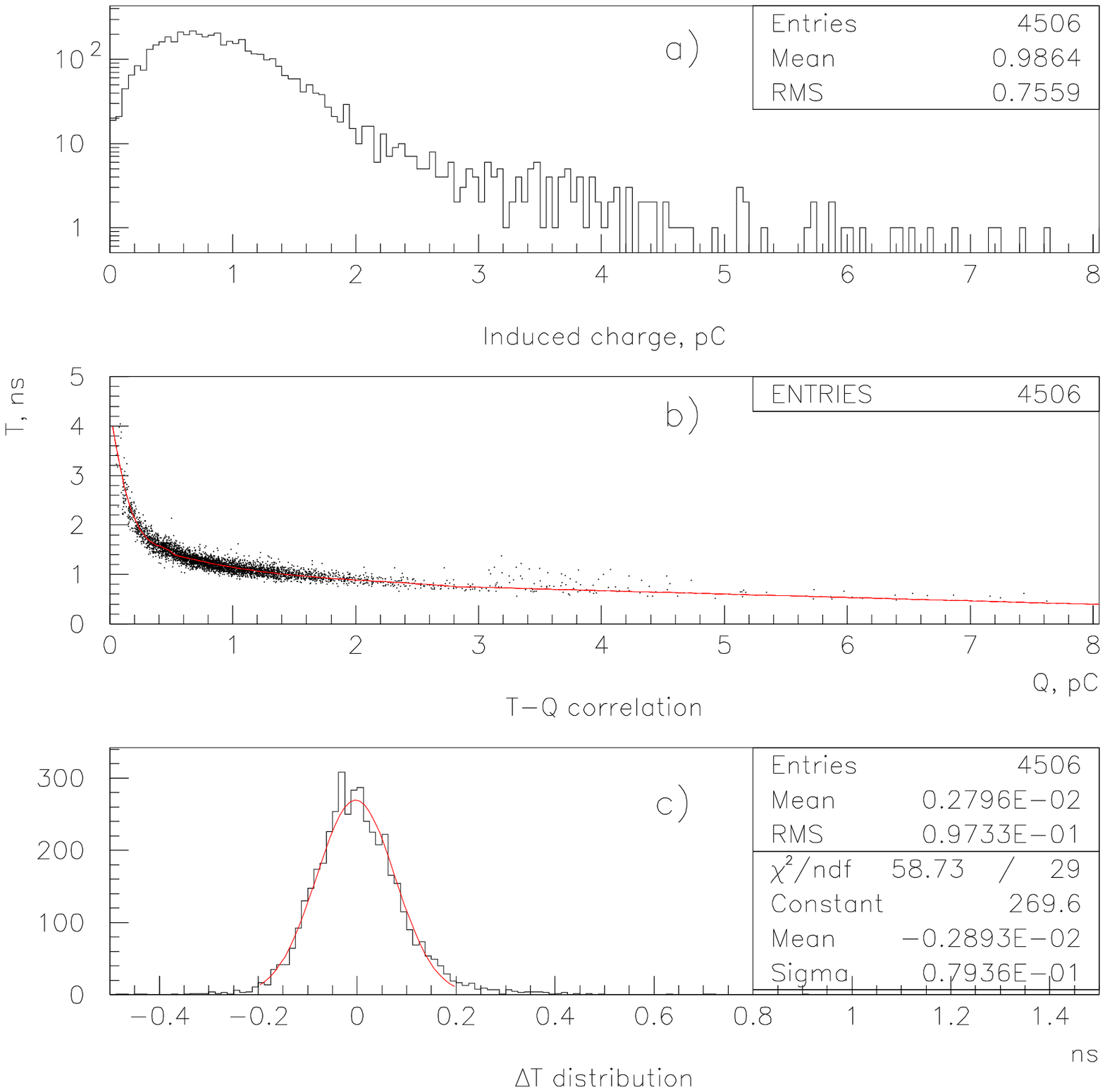}
\vfill
}
\vskip 10.5cm
\begin{minipage}[t]{17cm}
\begin{center}
\bf{Fig. 3.} 
\hskip 0.5cm
{Charge spectrum, arrival time - charge correlation plot and 
corrected time distribution for the 0.3 mm subgap counter. HV=6.4 kV} 
\end{center}
\end{minipage}
\end{figure}

\begin{figure}[hbt] 
\vbox to15cm{
\vfill
\includegraphics{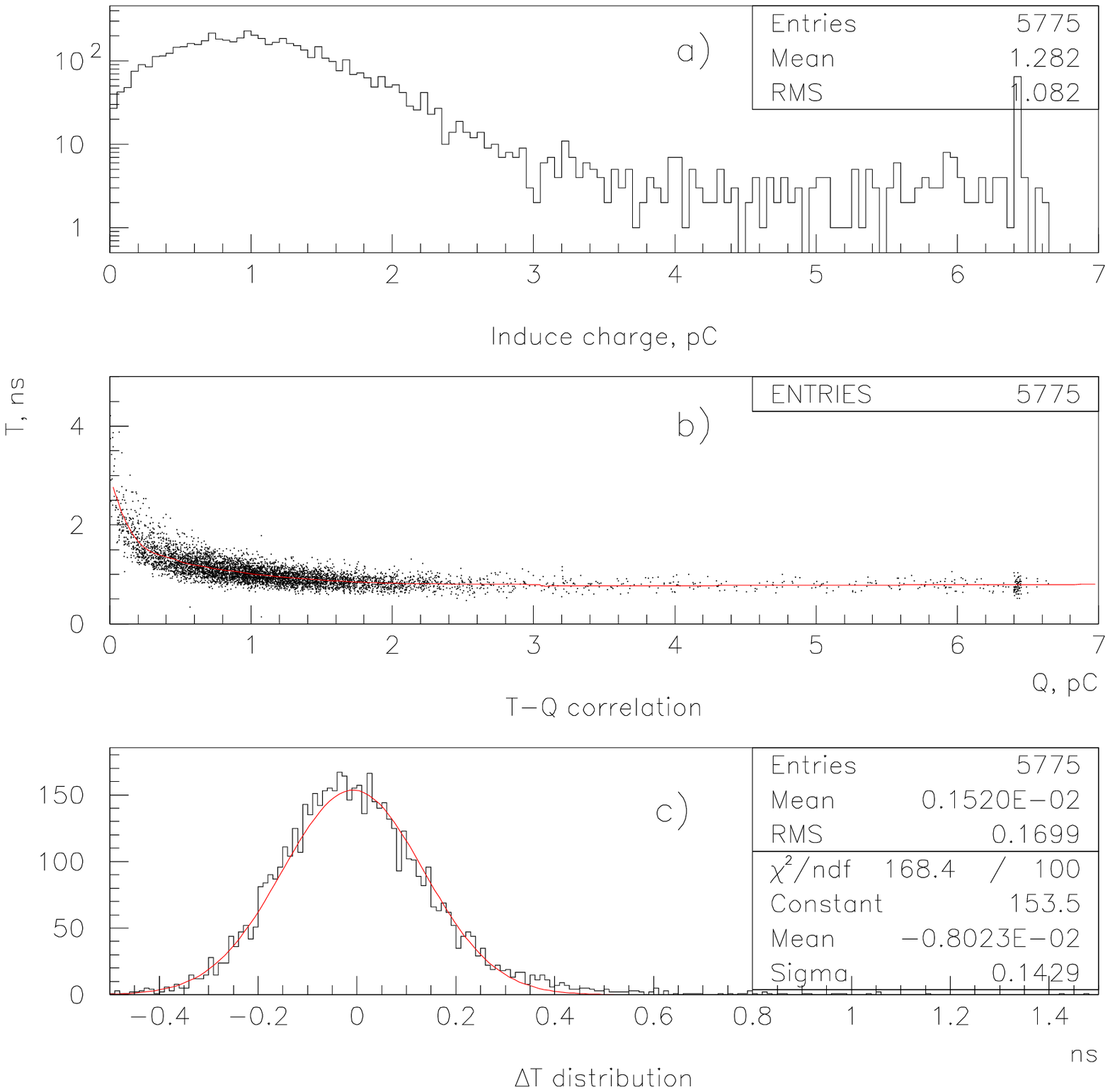}
\vfill
}
\vskip 7.5cm
\begin{minipage}[t]{16cm}
\begin{center}
\bf{Fig. 4.} 
\hskip 0.5cm
{Same as Fig. 3, for the 0.6 mm subgap counter. HV=5.4 kV.} 
\end{center}
\end{minipage}
\end{figure}

\begin{figure}[hbt] 
\vbox to15cm{
\vfill
\includegraphics{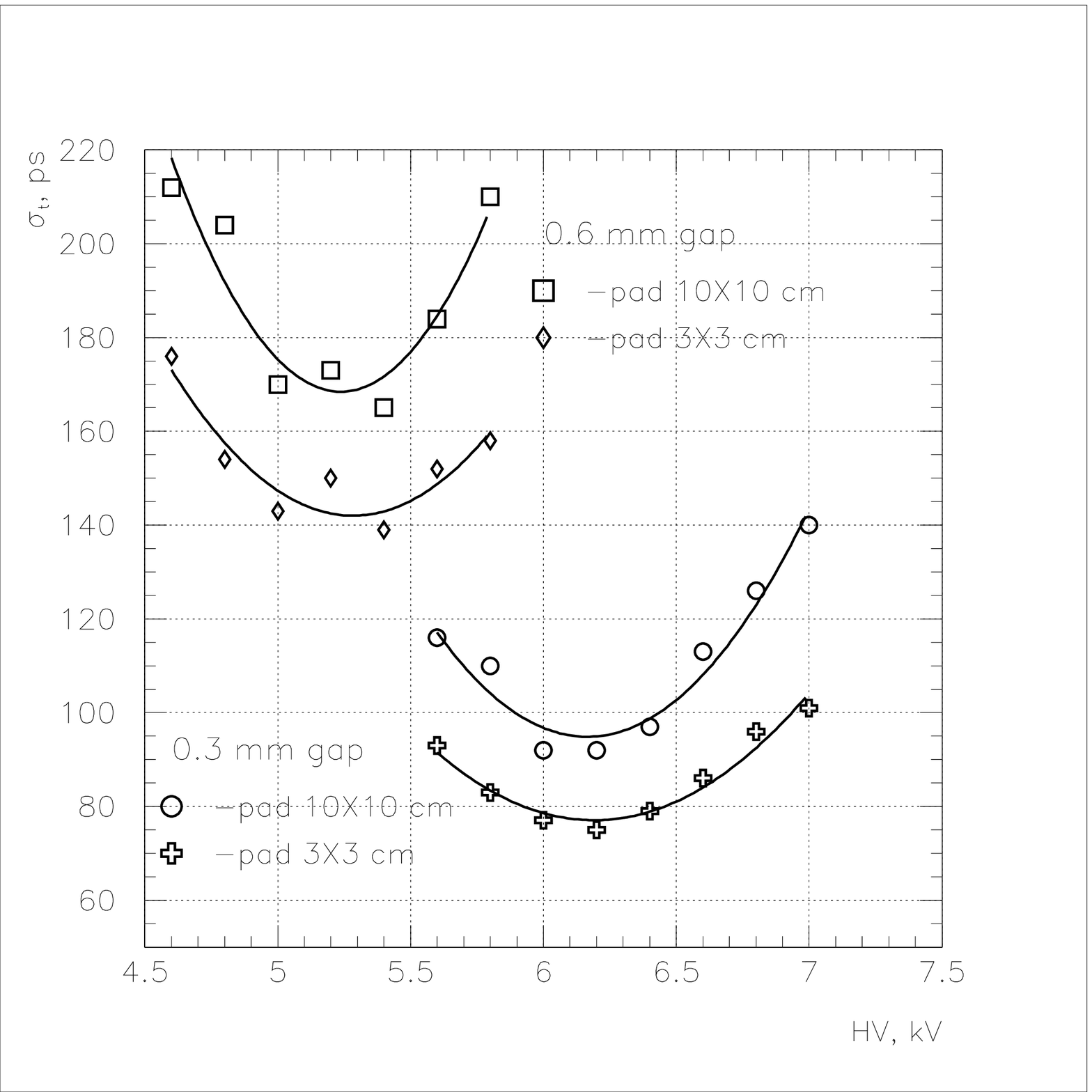}
\vfill
}
\vskip 7.5cm
\begin{minipage}[t]{16cm}
\begin{center}
\bf{Fig. 5.} 
\hskip 0.5cm
{Measured time resolution for two counters in dependence on applied HV.} 
\end{center}
\end{minipage}
\end{figure}

\begin{figure}[hbt] 
\vbox to15cm{
\vfill
\includegraphics{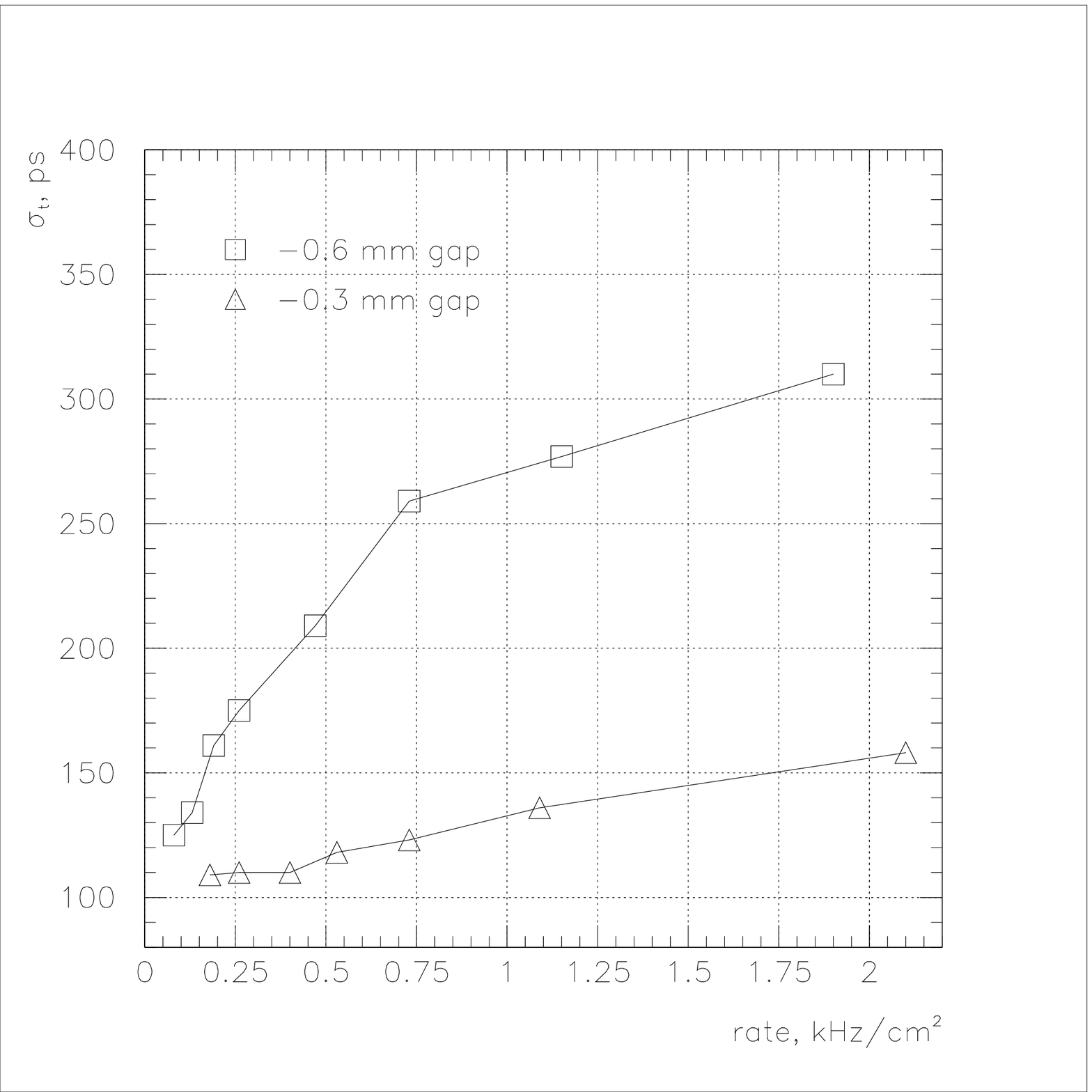}
\vfill
}
\vskip 7.5cm
\begin{minipage}[t]{16cm}
\begin{center}
\bf{Fig. 6.} 
\hskip 0.5cm
{Time resolution of two counters in dependence on counting rate.}
\end{center}
\end{minipage}
\end{figure}

\begin{figure}[hbt] 
\vbox to15cm{
\vfill
\includegraphics{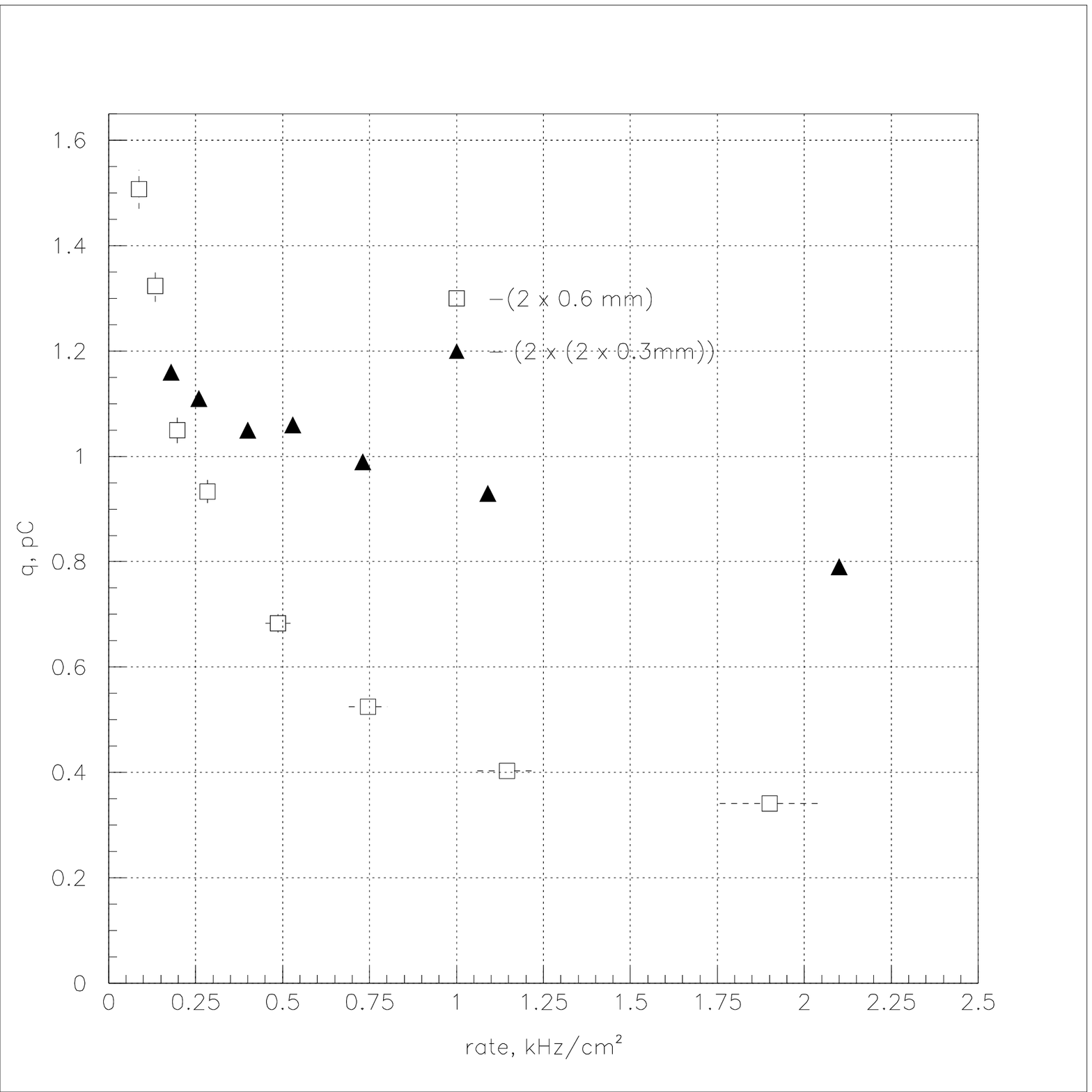}
\vfill
}
\vskip 7.5cm
\begin{minipage}[t]{16cm}
\begin{center}
\bf{Fig. 7.} 
\hskip 0.5cm
{Average charge dependence for two counters on counting rate.}
\end{center}
\end{minipage}
\end{figure}

\end{document}